\documentstyle[11pt,fleqn]{article}
\let\section\subsection

\begin{document}
\begin{flushright}
hep-th/9310037 \hfill UCLA/93/TEP/37 \\
October 1993
\end{flushright}
\vspace{11pt}
\begin{center}
{\Large 2-D GRAVITY AS GAUGE THEORIES WITH \\[3pt]
 EXTENDED GROUPS\footnote{Based on a talk given at the XXIIIth International
Conference on Differential Geometric Methods in Theoretical Physics. Ixtapa,
Mexico. September 1993.}} \\[11pt]
{\large DANIEL CANGEMI} \\[4pt]
{\it Department of Physics \\
University of California at Los Angeles \\
405 Hilgard Avenue \\
Los Angeles, CA 90024 (U.S.A.) } \\
\end{center}
\begin{abstract}
The interaction of matter with gravity in two dimensional spacetimes can be
supplemented with a geometrical force analogous to a Lorentz force produced
on a surface by a constant perpendicular magnetic field. In the
special case of constant curvature, the relevant
symmetry does not lead to the de Sitter or the Poincar\'e algebra but to
an extension of them by a central element. This richer structure suggests
to construct a gauge theory of 2-D gravity that reproduces the
Jackiw-Teitelboim model and the string inspired model. Moreover matter
can be coupled in a gauge invariant fashion. Classical and quantized results
are discussed.
\end{abstract}

\newcommand{\ie}{{\it i.e.,\ }}
\newcommand{\A}{{\cal A}}
\newcommand{\B}{{\cal B}}
\newcommand{\BB}{{\cal B}_\Lambda}
\newcommand{\s}[1]{{\textstyle{#1}}}
\newcommand{\diag}{\mathop{\rm diag}}
\newcommand{\id}{{\mathchoice {\rm 1\mskip-4mu l} {\rm 1\mskip-4mu l}
 {\rm 1\mskip-4.5mu l} {\rm 1\mskip-5mu l}}}
\newcommand{\centereqno}{\nonumber\\[-5.5pt] \\[-5.5pt]}

\section*{Introduction}

The beautiful success of General Relativity  and the key role played by gauge
theories in the description of fundamental interactions are two main reasons
leading physicists to be interested in differential geometry. On the one hand,
particles follow geodesics of spacetime,
on the other hand, gauge potentials are identified with connections on some
principal bundle. Moreover, it is tempting to exploit the local symmetries of
General Relativity to write it as a gauge theory. Attempts in this
direction turn out to be rather successful in lower dimensional gravities.
In 2+1 dimensions, it is recognized~\cite{townsend} that planar gravity is
described by a Chern-Simons model. In this note, I will consider the even
simpler case of 1+1 dimensions, where a gauge theoretical
formulation of lineal gravity has a natural setting using an
extended~\cite{can_jack}
Poincar\'e~\cite{verlinde} group or, more generally, an
extended~\cite{kim,can_dun} de Sitter~\cite{fukiyama,jackiw} group; the
extension is related to a geometrical force~\cite{force}, which exists only in
that particular dimension.

\section*{Gravity in 1+1 dimensions}

The reduction of General Relativity to 1+1 dimensions is not straightforward
because of the vanishing of the Einstein tensor. There are two main proposals
for lineal gravities.

One is obtained with a dimensional reduction of the Einstein-Hilbert action in
2+1 dimensions~\cite{teitelboim}.
\begin{equation}
I_{\rm JT} = {1\over2\pi k} \int d^2x \sqrt{-g} \eta (R - \Lambda)
\label{JT}
\end {equation}
The Lagrange multiplier $\eta$ enforces constant curvature, $R = \Lambda$.

The other proposal~\cite{CGHS} is inspired by string theory on a two
dimensional target space (it can alternatively be viewed as an s-wave
approximation of 3+1 gravity~\cite{harvey}).
\begin{equation}
\bar I_{\rm SI} = {1\over2\pi k} \int d^2x \sqrt{-\bar g} e^{-2\phi} (\bar R +
4 \bar g^{\mu\nu} \partial_\mu \phi \partial_\nu \phi - \lambda)
\label{liouville}
\end {equation}
Its classical solutions are $\bar g_{\mu\nu} = h_{\mu\nu} / (M - \lambda (x -
\bar x)^2)$, where $h_{\mu\nu} = \diag(1,-1)$ is the flat spacetime metric. The
value $M=0$ corresponds to a flat metric (vacuum solution), whereas the cases
$M \neq 0$ have the characteristics of a black hole. The
action~(\ref{liouville}) takes a simpler form with a change of
variables~\cite{verlinde}, $g_{\mu\nu} = \exp(-2\phi) \bar g_{\mu\nu}$, $\eta =
\exp(-2\phi)$.
\begin{equation}
I_{\rm SI} = {1\over2\pi k} \int d^2x \sqrt{-g} (\eta R - \lambda)
\label{SI}
\end {equation}
The Lagrange multiplier, $\eta$, now enforces zero curvature, $R = 0$.
Proposals~(\ref{JT}) and~(\ref{liouville}) suggest the more general
action~\cite{kim,can_dun}
\begin{equation}
I _{\rm g}= {1\over2\pi k} \int d^2x \sqrt{-g} \biggl(\eta (R - \Lambda) -
\lambda \biggr)
\label{action}
\end {equation}
In view of the string inspired model~(\ref{liouville}), the ``stringy'' metric
$\bar g_{\mu\nu}$ is conformally related to $g_{\mu\nu}$, $\bar g_{\mu\nu} =
g_{\mu\nu} / \eta$. However, there is no definite reason to prefer one or the
other as the physical metric~\cite{kubo}.

Let us end this section by recalling an equivalent formulation of geometry
where $(g_{\mu\nu}, R)$ is substituted with $(e^a_\mu, \omega_\mu)$. The {\it
Zweibein\/}, $e^a_\mu$, is related to the metric, $g_{\mu\nu} = e^a_\mu h_{ab}
e^b_\nu$, and the spin-connection, $\omega_\mu$, to the curvature, $d\omega =
R\,{\rm vol}/2$ (${\rm vol}$ is the volume two-form). Moreover, a space without
torsion implies a relation between the {\it Zweibein} and the spin-connection,
$de^a + \epsilon^a{}_b \omega e^b = 0$ ($\epsilon_{ab}$ is the antisymmetric
two-tensor with value $\epsilon^{01} = 1$).

\section*{Point particle motion on the line}

The gauge symmetry hidden in the action~(\ref{action}) becomes obvious if one
studies the motion of a particle on the line. The interaction of a point
particle in a background geometry is usually
described by the geodesic equation. However, in two dimensions (and only in
this dimension), the right side of that equation may be supplemented by a
force term of a geometrical nature~\cite{force}.
\begin{equation}
{d\over d\tau} {m \; \dot x^\mu\over\sqrt{\dot x^\alpha g_{\alpha\beta} \dot
x^\beta}} + {1\over\sqrt{\dot x^\alpha g_{\alpha\beta} \dot x^\beta}}
\Gamma^\mu_{\nu\rho} \dot x^\nu \dot x^\rho = {\cal F}(R)
g^{\mu\nu} \sqrt{-g} \epsilon_{\nu\rho} \dot x^\rho
\label{geodesic}
\end{equation}
This equation is still general covariant and invariant under reparametrization
provided ${\cal F}(R)$ is a scalar function. We will restrict ourself to linear
examples, ${\cal F}(R) = - \B - \A R / 2$. Due to its
similarity with electromagnetism (which is {\it not} included here),
the generalized geodesic
equation~(\ref{geodesic}) is obtained from the variation of the action,
\begin{eqnarray}
I_m = -  \int d\tau &\biggl[& m\, \sqrt{\dot x^\mu (\tau)
g_{\mu\nu}(x(\tau)) \dot x^\nu (\tau)} \nonumber\\
&& + \dot x^\mu (\tau) \biggl( \A
\omega_\mu (x(\tau)) + \B a_\mu (x(\tau)) \biggr) \biggr]
\label{matter}
\end{eqnarray}
where $\omega$ is the spin-connection and $a$ a one-form satisfying the
exactness condition $da =
{\rm vol}$.

It is easy to check that for constant curvature this action is
invariant under a change of coordinates defined by a Killing vector field.
Constant curvature spacetimes (with trivial topology, which we assume here) are
maximally symmetric and thus possess three independent Killing vectors fields,
$\xi^\mu_{(J)},
\xi^\mu_{(0)}, \xi^\mu_{(1)}$. By Noether's theorem, they generate three
conserved currents.
\begin{eqnarray}
\label{killing}
\xi^\mu_{(J)} = \epsilon^\mu{}_\nu x^\nu & \longrightarrow & J \centereqno
\xi^\mu_{(a)} = \delta^\mu_a (1-\s{\Lambda\over8} x^2) + \s{\Lambda\over4}
h_{a\nu} x^\nu x^\mu & \longrightarrow & P_a \hspace{33pt} (a=0,1) \nonumber
\end{eqnarray}
With the canonical symplectic structure $\left[ {\delta L\over \delta\dot
x^\mu} , x^\nu \right] = \delta^\nu_\mu$, these currents fulfill the
algebra,
\begin{eqnarray}
[P_a , J] &=& \epsilon_a{}^b P_b \centereqno
[P_a , P_b] &=& \epsilon_{ab} \bigl(\s{\Lambda\over2} J + \BB I \bigr)
\hspace{33pt} (\BB \equiv \B + \s{1\over2} \A \Lambda) \nonumber
\label{desitter}
\end{eqnarray}
where $I$ is a central element acting by 1 in the
representation~(\ref{killing}).

Due to the presence of a geometrical force, we do not  get the de Sitter
algebra in its expected form; more specifically, in the flat case, $\Lambda
= 0$, we do not recover the Poincar\'e algebra but a central extension of
it. For $\B \neq 0$, this algebra possesses a non-degenerate, invariant inner
product,
\begin{equation}
h_{AB} = \langle Q_A , Q_B \rangle = \left[
\begin{array}{ccc} h_{ab} & 0 & 0 \\
  0 & {(m/\BB)^2 \over 1 - {\Lambda\over2} (m/\BB)^2 } &
      - {1/\BB \over 1 - {\Lambda\over2} (m/\BB)^2 } \\
  0 & - {1/\BB \over 1 - {\Lambda\over2} (m/\BB)^2 } &
      {\Lambda/2\BB^2 \over 1 - {\Lambda\over2} (m/\BB)^2 } \end{array}
 \right]
\label{inner}
\end{equation}
($A,B = 0,1,2,3$; $Q_a = P_a$; $Q_2 = J$; $Q_3 = I$),
which depends on a real parameter $m$. The Casimir $Q_A h^{AB} Q_B$ in the
representation~(\ref{killing}) coincides with the Hamiltonian for a particle of
mass $m$. It can be shown that the freedom in the parameter $m$ corresponds in
the case $\Lambda = 0$ to a global  symmetry~\cite{jackiw} also found in the
dilaton model~\cite{russo} where its anomaly plays a crucial role in the
existence of Hawking radiation~\cite{kubo}.

\section*{Gauge formulation of the gravity sector}

We suggest to use this enhanced group structure for a gauge description of
gravity. A connection will be thus a one-form of the type
\begin{equation}
A = e^a P_a + \omega J + \BB a I
\end{equation}
with curvature two-form
\begin{eqnarray}
F &=& dA + A^2 \\
  &=& (de^a + \epsilon^a{}_b \omega e^b) P_a + (d\omega + \s{\Lambda\over4}
e^a \epsilon_{ab} e^b) J + \BB (da + \s{1\over2} e^a \epsilon_{ab} e^b) I
\nonumber
\end{eqnarray}
The components of $F$ reproduce geometrical quantities if we interpret $e^a$ as
a {\it Zweibein} and $\omega$ as a spin-connection: The two first components
are the torsion relating the {\it Zweibein} to the spin-connection, the third
one equals $(R - \Lambda) {\rm vol} /2$ and the last one $(da - {\rm vol})$.
Using a scalar function with value in the adjoint representation of
the gauge group, $\eta = \eta^a P_a + \eta^2 J + \eta^3 I$,
and the non-degenerate inner product~(\ref{inner}), we build a gauge invariant
action,
\begin{eqnarray}
I'_{\rm g} &=& {1\over2\pi k} \int \langle \eta, F \rangle \nonumber\\
&=& {1\over2\pi k} \int \Bigl[ \eta_a (de^a + \epsilon^a{}_b \omega e^b)
\nonumber\\
&&\hspace{14mm} - \s{1 \over 1 - {\Lambda\over2} (m/\BB)^2} \Bigl((m/\BB)^2
\eta^2 - \s{1\over\BB} \eta^3\Bigr) (d\omega + \s{\Lambda\over4} e^a
\epsilon_{ab} e^b) \nonumber \\
&&\hspace{14mm} + \s{1 \over 1 - {\Lambda\over2} (m/\BB)^2} \Bigl(- (\eta^2 +
\s{\Lambda\over2\BB} \eta^3\Bigr) (da + \s{1\over2} e^a \epsilon_{ab} e^b)
\Bigr]
\end{eqnarray}
which not only reproduces the action~(\ref{action}) with
\begin{eqnarray}
\eta &=& \s{1\over1 - {\Lambda\over2} (m/\BB)^2} \left( \s{1\over2} (m/\BB)^2
\eta^2 - \s{1\over2\BB} \eta^3 \right) \centereqno
\lambda &=& \s{1\over1 - {\Lambda\over2} (m/\BB)^2} \left( - \eta^2 +
\s{\Lambda\over2\BB} \eta^3 \right) \nonumber
\end{eqnarray}
but also provides a one-form, whose classical value, $da = {\rm vol}$, is the
one needed to construct the matter action~(\ref{matter}).

Besides the zero curvature condition, $F = 0$, we also get an equation for
the scalar function,
$D_\mu \eta = 0$. This set of equations is easily solved by the general
solution
\begin{eqnarray}
A &=& U^{-1} d U \centereqno
\eta &=& U^{-1} \eta_{(0)} U \nonumber
\end{eqnarray}
for any group element $U$ and constant gauge algebra element $\eta_{(0)}$.
Of course, $U$ has to be chosen carefully in order to reproduce a
geometric solution associated to a non-degenerate metric
$g_{\mu\nu}$~\cite{can_dun,jackiw}. The ``stringy'' metric $\bar g_{\mu\nu} =
g_{\mu\nu} / \eta$ then takes the form of a static black hole, for $\Lambda =
0$, $\bar g_{\mu\nu} = h_{\mu\nu} / (M - \lambda (x - \bar x)^2)$. Nevertheless
the physical content of the model will not depend on this choice and $U =
\id$ \ie $e^a = \omega = a = 0$, is perfectly admissible. This is
sometimes referred as the unbroken phase.  The physics should be contained in
the gauge invariant part of $\eta$.
\begin{eqnarray}
&& \langle \eta, \eta \rangle = \langle \eta_{(0)}, \eta_{(0)} \rangle = M
\centereqno
&& \langle \eta, I \rangle = \langle \eta_{(0)}, I \rangle = \lambda/\BB
\nonumber
\end{eqnarray}

The gauge theoretical approach relates the number of free parameters in the
classical solutions $(M, \lambda, \bar x^0, \bar x^1)$ to the dimension (four)
of the gauge group. It introduces also the cosmological constant $\lambda$ as a
dynamical variable. The parameters $M$ and $\lambda$ are gauge invariant
quantities and describe the physical content of the theory, as we will see in
the next section.

\section*{Quantization of the gravity sector}

A gauge theoretical setting allows a more tractable way to deal with
quantization. We present here the canonical quantum structure
of gravity without matter; it is simple and interesting, even if, in the
absence of matter, there are no propagating degrees of freedom. We write
the action~(\ref{action}) in its Hamiltonian form.
\begin{eqnarray}
I'_g &=& {1\over2\pi k} \int d^2x \epsilon^{\mu\nu} \langle
\eta,F_{\mu\nu} \rangle \\
&=& {1\over2\pi k} \int dt dx \Bigl( \langle \eta,\partial_0 A_1 \rangle +
\langle A_0,D_1 \eta \rangle \Bigr) - {1\over2\pi k} \int dt dx \partial_1
\langle \eta,A_0 \rangle \nonumber
\end{eqnarray}
The Hamiltonian is a sum of constraints
\begin{equation}
G^A = - (\partial_1 \eta^A + f_{BC}{}^A A_1^B \eta^C)
\end{equation}
($A,B,C=0,1,2,3$ are the gauge group indices, which are raised and lowered with
the inner product $h_{AB}$, and $f_{BC}{}^A$ are the structure
constants of the gauge group). The spatial component of the gauge connection is
canonically conjugate to $\eta$ and we postulate the usual commutation
relations
\begin{equation}
[ \eta_A(x), A_1^B(y)] = i\, 2\pi k\, \delta_A^B \delta(x-y)
\end{equation}
With these commutation relations, the algebra of constraints coincides, as
usual in gauge theories, with the original gauge algebra.
\begin{equation}
[G_A(x), G_B(y)] = i f_{AB}{}^C G_C(x) \delta(x-y)
\end{equation}

In a Schroedinger picture, we consider states as functionals of $\eta_A(x)$,
$\Psi[\eta_A]$, on which $A_1^A(x)$ acts by functional derivation, $(2 \pi k /
i) (\delta / \delta \eta_A(x))$. Physical states are those annihilated by
the constraints $G_A$ and they satisfy the differential equations
\begin{eqnarray}
 \biggl( \partial_1 \eta_a - i\, 2\pi k\, \epsilon_a{}^b \eta_b
{\delta\over\delta\eta^2} + i\, 2\pi k\, \eta^2 \epsilon_{ab}
{\delta\over\delta\eta_b} \biggr) \Psi &=& 0 \nonumber \\
 \biggl( \partial_1 \eta^2 + i\, 2\pi k\, {\Lambda\over4} \epsilon^a{}_b
\eta_a {\delta\over\delta \eta_b} \biggr) \Psi &=& 0 \\
 \biggl( \partial_1 \eta^3 + i\, 2\pi k\, \BB \epsilon^a{}_b \eta_a
{\delta\over\delta \eta_b} \biggr) \Psi &=& 0 \nonumber
\end{eqnarray}
These equations are solved by the functionals
\begin{equation}
\Psi[\eta_A] = \left. \exp \left( {i\over2\pi k} \int dx\, \eta^2
{\epsilon^{ab} \partial_1 \eta_a \eta_b\over \eta_c \eta^c} \right) \psi
(M,\lambda) \right|_{\langle \eta, \eta \rangle = M \atop \langle \eta, I
\rangle = \lambda/\BB}
\end{equation}
with support on the constant gauge invariant combinations $\langle \eta,\eta
\rangle = M$ and $\langle I, \eta \rangle = \lambda / \BB$; $\psi$ is a
function of the variables $M$ and $\lambda$. The physical states depend on the
two values $M$ and $\lambda$, which coincide for classical solutions with the
two parameters of the black hole configuration. Let us now couple matter to
this gravity.

\section*{Coupling to matter}

The coupling to matter follows the one discussed before, see
Eq.~(\ref{matter}). It is possible to find a gauge invariant formulation of it
either for point particle or for fields, {\it cf.\/}~Ref.~\cite{force}. The
gauge invariant actions are of the form
\begin{equation}
I'_m[A_\mu,p(\tau),\xi^a], \;\; I'_m[A_\mu,\bar\psi,\psi,\xi^a], \;\; \ldots
\end{equation}
where the additional field $\xi^a$ acts like a Higgs field that insures the
gauge invariance of the action. The essential feature of this coupling is that
it does not involve $\eta$. In this gauge formulation, the matter is coupled to
the metric $g_{\mu\nu}$, whereas in the geometrical point of view people use
mainly a coupling to $\bar g_{\mu\nu}$. But, since their coupling is conformal,
it is not really different at the classical level. Nevertheless, this
difference could have its importance once we proceed to the
quantization~\cite{kubo}. Notice that our coupling breaks conformal invariance
at the classical level even in the massless case. Namely, the trace of the
energy-momentum is proportional to the additional force strength, $\B$ and at
the quantum level its vacuum expectation picks up an additional term,
$R/24\pi$~\cite{force}.

The equations of motion are modified in the following way
\begin{eqnarray}
F &=& 0 \centereqno
D_\mu \eta &=& 2\pi k \, J^5_\mu \nonumber
\end{eqnarray}
where $(J^5_\mu)_A = - \epsilon_{\mu\nu} (\delta I'_m / \delta A^A_\nu)$ is the
axial current. Let us consider the point particle. Outside the particle
trajectory, $J^5_\mu$ is zero and the equations are those of pure gravity. We
have two sets of four constant parameters on each side of the trajectory, whose
differences are fixed by the particle characteristics. The shift in $M$ and
$\lambda$ implies a transition from a pure gravity state to another when
crossing the particle line; this is usually interpreted~\cite{CGHS} as a black
hole created by an in-falling particle. The shift in $\bar x$ is a basic
ingredient in deriving a Hawking radiation~\cite{CGHS} for the ``stringy''
metric, $\bar g_{\mu\nu}$.

Our formulation reproduces interesting features of lineal gravity. But being  a
gauge theory, we are able to discuss in a straightforward manner issues
concerning gauge charges or quantization.

\section*{A gauge definition of mass}

The definition of mass and angular-momentum is an ill-defined concept in
General Relativity. Different methods lead to different results~\cite{bak}.
However, when one has a gauge invariance, Noether's procedure uniquely define
conserved currents and charges. In our model, $I'_g + I'_m$, an infinitesimal
gauge transformation $\theta$ generates an explicit conserved current
\begin{equation}
j^\mu_\theta = {1\over\pi k} \epsilon^{\mu\nu} \partial_\nu \langle \eta,
\theta \rangle
\end{equation}
and a conserved charge.
\begin{equation}
Q_\theta = \int dx^1 j^0_\theta = {1\over\pi k} \langle \eta, \theta \rangle
\bigr|^{x^1=+\infty}_{x^1=-\infty}
\end{equation}
The question is which $\theta$ define energy. Obviously, energy should be
related with infinitesimal diffeomorphisms in a time-like Killing direction.

But, in topological field theory ($F=0$), infinitesimal diffeomorphisms are
equivalent to infinitesimal gauge transformations~\cite{lie}.
\begin{equation}
L_f A_\mu = f^\alpha \partial_\alpha A_\mu + \partial_\mu f^\alpha A_\alpha =
D_\mu (f^\alpha A_\alpha) + f^\alpha F_{\alpha\mu}
\end{equation}
An infinitesimal diffeomorphism, $f^\alpha$, is identified with an
infinitesimal gauge transformation, $f^\alpha A_\alpha$. It is thus associated
to the conserved charge
\begin{equation}
Q_f = {1\over\pi k} \langle \eta, f^\alpha A_\alpha \rangle
\bigr|^{x^1=+\infty}_{x^1=-\infty}
\end{equation}
and energy $E$ is defined for a time-like Killing vector $f^\alpha$.

In the absence of matter, the contributions at $x^1=+\infty$ and $x^1=-\infty$
are identical, which implies $E=0$. When matter is included, due to the jump of
the value of $\eta$ across the particle trajectory, the contributions are
different and gives a non zero energy, $E=\langle \eta, \eta \rangle = M$, in
full agreement with the ADM definition.

\section*{Conclusions}

In this brief note, I have shown how General Relativity and gauge theory can be
combined in 1+1 dimensional spacetime. Once the gauge group is recognized, we
are able to produce a gauge theory, which encompasses the Jackiw-Teitelboim and
the string inspired models. The inclusion of matter in a gauge invariant way is
possible and provides a model, which not only reproduces previous results but
also provides a natural way to define gauge invariant and conserved quantities,
as energy, and to deal with quantization. Another interesting feature of the
model is the introduction of the cosmological constant as a dynamical
variable~\cite{izawa}. Supersymmetric extensions have been studied in relation
to a positive energy theorem~\cite{park} and for a topological description of
supergravity~\cite{martin}. The quantization of pure gravity has shown how the
physical states depend on gauge invariants. The quantization of the full model
deserves further study. It would also be interesting to consider topological
effects occuring in the definition of the one-form $a$ and in the
resolution of $F = 0$~\cite{hwang}.

\section*{Acknowledgements}

I thank G.~Dunne and R.~Jackiw for helpful comments.
This work is supported in part by funds provided by N.S.F. under contract
PHY-89-15286 and by the ``Fondation du 450e anniversaire de l'Universit\'e de
Lausanne.''


\begin{thebibliography}{}

\bibitem{townsend}
Ach\`ucarro, A. and Townsend, P.: 1986, {\it Phys. Lett.\/} {\bf 180B}, 89;
Witten, E.: 1988/89, {\it Nucl. Phys.\/} {\bf B311}, 46.

\bibitem{CGHS}
Callan, C., Giddings, S., Harvey, J. and Strominger, A.: 1992, {\it Phys.
Rev. D\/} {\bf 45}, 1005;
for a review, see Mann, R.: 1993, ed(s)., {\it Proceedings of the Fourth
Canadian Conference on General Relativity and Relativistic Astrophysics,\/} to
be published.

\bibitem{bak} Bak, D., Cangemi, D. and Jackiw, R.: 1993, `Energy-Momentum
Conservation in General Relativity', {\it MIT preprint} {\bf CTP\#2245}.

\bibitem{can_dun} Cangemi, D. and Dunne, G.: 1993, `Extended de Sitter Theory
of
Two-Dimensional Gravitational Forces', to appear in {\it Phys. Rev. D\/}.

\bibitem{can_jack} Cangemi, D. and Jackiw, R.: 1992, {\it Phys. Rev. Lett.\/}
{\bf 69}, 233.

\bibitem{force} Cangemi, D. and Jackiw, R.: 1993, {\it Phys. Lett. B\/} {\bf
299}, 24;
 Cangemi, D. and Jackiw, R.:1993, {\it Ann. Phys. (NY)\/} {\bf 225}, 229.

\bibitem{martin} Cangemi, D. and Leblanc, M.: 1993, `Two Dimensional Gauge
Theoretic Supergravities', {\it MIT preprint} {\bf CTP\#2224}.

\bibitem{kubo} Fujiwara, T., Igarashi, Y., Kubo, J.: 1993, `Unrecognizable
Black Holes in Two Dimensions', {\it Kanazawa preprint} {\bf 93-01} and `Weyl
Invariance and Spurious Black Hole in Two-Dimensional Dilaton Gravity',
{\it Kanazawa preprint} {\bf 93-06}.

\bibitem{fukiyama} Fukiyama, T. and Kamimura, K.: 1985, {\it Phys. Lett. B\/}
{\bf 160}, 259;
Isler, K. and Trugenberger, C.: 1989, {\it Phys. Rev. Lett.\/} {\bf 63}, 834;
Chamseddine, A. and Wyler, D.: 1989, {\it Phys. Lett. B\/} {\bf 228}, 75.

\bibitem{harvey} Harvey, J. and Strominger, A.: 1992, in Quantum Aspects of
Black Holes, ed(s)., {\it String Theory and Quantum Gravity '92\/}, Spring
School and Workshop: Trieste, to be published.

\bibitem{hwang} See for example Hwang, D., Kim, S., Soh, K. and Yee, J.: 1993,
`Particle Motion in a Curved Cylindrical Geometry', to appear in {\it Phys.
Rev. D\/};
Iengo, R. and Li, D.: 1993, `Quantum Mechanics and Quantum Hall Effect on
Riemann Surfaces', {\it SISSA Preprint} {\bf 100/93}.

\bibitem{izawa} Izawa, K.: 1993, `Dynamics of the Cosmological Constant in
Two-Dimensional Universe', {\it Kyoto preprint} {\bf KUNS 1221}.

\bibitem{lie} Jackiw, R.: 1978, {\it Phys. Rev. Lett.\/} {\bf 41}, 1635.

\bibitem{jackiw} Jackiw, R.: 1992, {\it Theor. Math. Phys.\/} {\bf 9}, 404.

\bibitem{kim} Kim, S., Soh, K. and Yee, J.: 1993, {\it Phys. Rev. D\/} {\bf
47},
4433.

\bibitem{park} Park, Y. and Strominger, A.: 1993, {\it Phys. Rev. D\/} {\bf
47},
1569.

\bibitem{russo} Russo, J., Susskind, L. and Thorlacius, L.: 1992, {\it Phys.
Rev. D\/} {\bf 46}, 3444.

\bibitem{teitelboim} Teitelboim, C.: 1983, {\it Phys. Lett. B\/} {\bf 126}, 41
and 1984, ed(s)., {\it Quantum Theory of Gravity\/}, S. Christensen, Adam
Hilger: Bristol;
Jackiw, R.: 1984, ed(s)., {\it Quantum Theory of Gravity\/}, S. Christensen,
Adam
Hilger: Bristol and 1985, {\it Nucl. Phys. B\/} {\bf 252}, 343.

\bibitem{verlinde} Verlinde, H.: 1992, ed(s)., {\it Sixth Marcel Grossmann
Meeting on General Relativity}, M.~Sato, World Scientific: Singapore;
Grignani, G. and Nardelli, G.: 1993, `Poincar\'e Gauge Theories for Lineal
Gravities', to appear in {\it Nucl. Phys. B\/}.

\end{thebibliography}
\end{document}